\documentclass[aps, prb, floatfix, reprint]{revtex4-2}
\usepackage{graphicx}
\usepackage{dcolumn}
\usepackage{bm}
\usepackage{amsfonts}
\usepackage{verbatim}
\usepackage{amsmath}
\usepackage{amssymb}
\usepackage{color}
\usepackage[normalem]{ulem}
\usepackage[percent]{overpic}
\usepackage{mathtools}
\usepackage{tikz}
\usepackage[colorlinks=true,citecolor=blue]{hyperref}
\hypersetup{colorlinks=true,citecolor=blue,linkcolor=red,urlcolor=blue}
\newcommand{\sgn}{\operatorname{sgn}}

\def\be{\begin{equation}}
\def\ee{\end{equation}}

\def\rv{{\bf r}}
\def\Av{{\bf A}}
\def\Bv{{\bf B}}
\def\hv{{\bf h}}
\def\kv{{\bf k}}
  
\def\pv{{\bf p}}

\def\tauv{{\bm \tau}}

\begin{document}

\title{Quantum Hall effect and optical magneto-conductivity of two-dimensional topological nodal-line semimetals}
\author{Shahin Barati}
\affiliation{Department of Physics, Institute for Advanced Studies in Basic Sciences (IASBS), Zanjan 45137-66731, Iran}
\author{Hamid Rahimpoor}
\affiliation{Department of Physics, Institute for Advanced Studies in Basic Sciences (IASBS), Zanjan 45137-66731, Iran}
\author{Saeed H. Abedinpour}
\email{abedinpour@iasbs.ac.ir}
\affiliation{Department of Physics, Institute for Advanced Studies in Basic Sciences (IASBS), Zanjan 45137-66731, Iran}
%\affiliation{Research Center for Basic Sciences \& Modern Technologies (RBST),  Institute for Advanced Studies in Basic Sciences (IASBS), Zanjan 45137-66731, Iran}
\date{\today}

\begin{abstract}
We investigate a gapped two-dimensional nodal-line semimetal subjected to a perpendicular magnetic field.
We identify an unusual pattern for Landau levels, where the energy of Landau levels first decreases and then starts increasing versus the level index. 
These descending/ascending Landau levels form a \emph{stretched check-mark} shape.
Tuning the magnitude of the magnetic field changes the spacing between Landau levels. 
At fixed chemical potential in the low doped regime, the Landau levels move below the chemical potential and get filled from one side, while they move out and get vacant from the other side. 
Using the Kubo formalism in the linear response theory, we analytically study a gapped nodal-line semimetal's longitudinal and Hall conductivities.
In the DC limit, we observe that as the magnetic field is varied at a fixed chemical potential, the Chern number swings back and forth between two consecutive integer numbers, giving rise to \emph{bumpy} Hall plateaus in a low-doped system.
We also find that the longitudinal optical conductivity displays different series of the Shubnikov-de Haas oscillations, corresponding to inter-band and intra-band transition among the descending and ascending sets of the Landau levels. 
\end{abstract}
\maketitle

%%%%%%%%%%%%%%%%%%%%%%%%%%%%%%%%%%%
\section{Introduction}\label{introduction}
Topological semimetals, materials in which conduction and valence bands touch at points or lines in the momentum space, have attracted immense interest lately. 
Nodal point semimetals such as Weyl~\cite{wan2011topological, lu2015experimental} and Dirac~\cite{wang2013three, borisenko2014experimental} semimetals have been intensively studied both
theoretically and experimentally~\cite{xu2015discovery, wang2013three,wan2011topological}. 
The bulk conduction and valence bands manifest linear dispersion near the discrete band-touching points in the Brillouin zone (BZ) for both Weyl and Dirac semimetals. 
In two dimensions (2D), graphene~\cite{neto2009electronic} exhibits such Dirac points between
the conduction and valence bands.

Nodal-line semimetals (NLSMs)~\cite{weng2015topological,fang2016topological} are another class of topological semimetals. 
In three spatial dimensions, they are characterized by band-touching lines or rings in the momentum space protected by various combinations of time reversal invariance, inversion, chiral, or other lattice symmetries.
Burkov \emph{et al.} first proposed an explicit model of three-dimensional NLSMs  in 
a superlattice of normal and topological insulators~\cite{burkov_prb2011}. 
Then, three-dimensional (3D) NLSMs have been theoretically proposed in many platforms. Cubic antiperovskite Cu$_3$PdN~\cite{yu2015topological}, hyperhoneycomb lattices~\cite{mullen2015line}, body-centered orthorhombic C$_{16}$~\cite{wang_prl2016}, families of CaP$_3$~\cite{Xu_PRB2017} and Ca$_3$P$_2$~\cite{Chan_PRB2016} structures, and TlTaSe$_2$~\cite{bian2016drumhead} are just a few examples.
Experimental evidence of nodal lines in PbTaSe$_2$~\cite{bian2016topological}, InBi~\cite{ekahana_njp2017},  and ZrSiS~\cite{schoop2016dirac}, as well as several other components through angle-resolved photoemission spectroscopy (ARPES), are reported. 

Similar to many other topological materials, NLSMs also have two-dimensional analogs. 
Lu \emph{et al.} proposed that a mixed lattice composed of Kagome and honeycomb structures can host 2D nodal-line states~\cite{lu2017two}. 
Utilizing a simple three-band tight-binding model for a Lieb lattice, Yang \emph{et. al.}, demonstrate that Be$_2$C and BeH$_2$ monolayers can host 2D nodal lines~\cite{Yang2017DiracNL}. They showed that structural fluctuations on the BN substrate open a trivial band gap in BeH$_2$ monolayer, but the node-line is robust against such perturbations in Be$_2$C. 
A family of two-dimensional node-line semimetals is predicted in MX (M=Pd, Pt; X=S, Se, Te) structures~\cite{Jin2017} through first-principles calculations.
Furthermore, H. Chen \emph{et al.} found that the nodal line in C$_9$N$_4$ forms a closed loop centered at $ \Gamma $ point in the BZ. Such a nodal loop originates from the p$_{\rm z}$ orbitals of both C and N atoms~\cite{Chen23}. 
Nodal lines in 2D are generally very fragile. Spin-orbit coupling or buckling may open a finite band gap in the band-touching line or convert it into a 2D Dirac semimetal~\cite{lu2017two}.

Various optical properties of topological semimetals have attracted much interest.
The AC conductivity of two-dimensional~\cite{PhysRevB.86.195405} and three-dimensional~\cite{PhysRevB.94.165111,PhysRevB.93.085426,PhysRevB.93.085442,PhysRevB.89.245121,PhysRevB.100.085436,PhysRevB.99.115406} Dirac and Weyl semimetals are studied. 
Optical conductivity of three and two-dimensional nodal line semimetals for clean~\cite{Carbottearticle} and dirty~\cite{PhysRevB.95.214203} and anisotropic models~\cite{shahin_prb2017} as well as the magneto-optical conductivity~\cite{PhysRevB.102.195123} and anomalous Hall optical conductivity~\cite{PhysRevB.103.165104} of NLSMs are also investigated. 

This work studies a gapped two-dimensional nodal-line semimetal subjected to a strong perpendicular magnetic field. 
Using the Kubo formula, we analytically calculate this system's AC and DC longitudinal and Hall conductivities at different carrier concentration regimes. The structure of the Landau levels of 2D NLSM and their magnetic field dependence suggests an intriguing integer quantum Hall effect in this system. 

The rest of this paper is organized as follows. 
Sec.~\ref{sec:model} introduces a low-energy model Hamiltonian for 2D NLSMs and discusses its Landau levels.
In Sec.~\ref{sec:kubo}, we analytically calculate the real and imaginary parts of longitudinal and Hall optical conductivity. The static limit and the integer quantum Hall effect are discussed in Sec.~\ref{sec:DC-Hall}.
In Sec.~\ref{sec:Gapless}, we study the optical conductivity of gapless 2D nodal-line semimetals.
Finally, we summarize our main findings in Sec.~\ref{sec:summary}.

%%%%%%%%%%%%%%%%%%%%%%%%%%%%%%%%%%%
\section{Model Hamiltonian and Landau levels}\label{sec:model}
\begin{figure*}
	\centering
	 \begin{tikzpicture}[scale=1.]
	 	\node at (0,0) {\includegraphics[width=\linewidth]{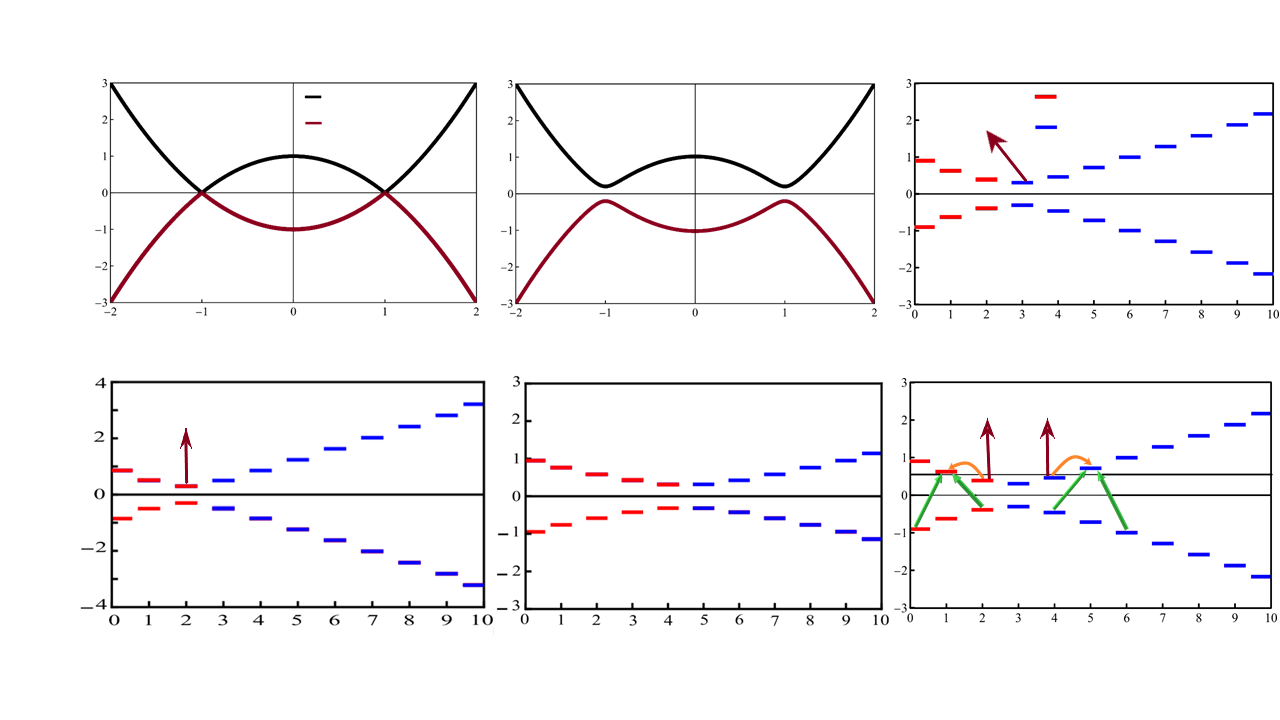}};
	 \draw (-8.1,2.5) node [font=\large][rotate=90] {$ {\varepsilon}_{k,s}/\varepsilon_r $};
	 \draw (-4.1,3.7) node [font=\large] {$ \varepsilon_{+} $};
	 \draw (-4.1,3.3) node [font=\large] {$ \varepsilon_{-} $};
	 \draw (-5.5,3.6) node [font=\large] {$ \Delta=0 $};
	 \draw (-6.8,3.6) node [font=\large] { (a) };
	 \draw (-4.8,0.15) node [font=\large] {$ k_x/k_r $};
	 \draw (0.8,0.15) node [font=\large] {$ k_x/k_r $};
	 \draw (0.1,3.58) node [font=\large] {$ \Delta\neq 0 $};
	 \draw (-1.1,3.58) node [font=\large] { (b) };
	 \draw (3.5,2.5) node [font=\large][rotate=90] {$ E_{n,s}/\varepsilon_r $};
	 \draw (6.5,0.15) node [font=\large] {$ n $};
	 \draw (-8,-2.1) node [font=\large][rotate=90] {$ E_{n,s}/\varepsilon_r $};
	 \draw (-4.8,-4.2) node [font=\large] {$ n $};
	 \draw (5,3.4) node {$ {\rm LLL} $};
	 \draw (-6.1,-0.75) node {$ {\rm LLL} $};
	 \draw (7.1,3.68) node {descending LLs};
	 \draw (7,3.3) node {ascending LLs};
	 \draw (4.2,3.6) node [font=\large] {(c) };
	 \draw (-7,-0.7) node [font=\large] { (d) };
	 \draw (-1.1,-0.7) node [font=\large] { (e) };
	 \draw (4.2,-0.7) node [font=\large] { (f) };
	 \draw (0.5,-4.2) node [font=\large] {$ n $};
 	 \draw (8.4,-1.35) node [font=\large]{${\mu}/{\varepsilon_r }$};
	 \draw (6.6,-4.2) node [font=\large] {$ n $};
	 \draw (5,-0.65) node [font=\large] {$ n^<_\mathrm{F} $};
	 \draw (5.85,-0.65) node [font=\large] {$ n^>_\mathrm{F} $};
	 	\end{tikzpicture}    
	\caption{Low-energy spectrum of a (a) gapless; and (b) gapped 2D NLSM versus $ k_x/k_r $ for $ k_y=0 $. 
	(c)-(e) Landau levels of a gapped 2D NLSM versus level index $n$ for (c) the generic case with $\varepsilon_r/\varepsilon_c= 3.3$; (d) for a special tunning of the magnetic field where $\varepsilon_r/\varepsilon_c = 2.5$  is half-integer;   and  (e) for the case where $\varepsilon_r/\varepsilon_c= 5$  is integer. 
	Descending and ascending LLs are identified by red and blue colors, respectively. The lowest Landau level (LLL) is also shown in panels (c) and (d).
	(f) An Illustration of LLs where a typical chemical potential $\mu$ for a low-doped system is shown together with the highest occupied levels from descending ($n^<_{\rm F}$) and ascending ($n^>_{\rm F}$) sets of levels. 
	Orange and green arrows mark the allowed intra-band and inter-band optical transitions.
	We have fixed the gap parameter to $\Delta=0.3\,\varepsilon_r $ in panels (b)-(f).
		\label{fig:LL}}
\end{figure*}
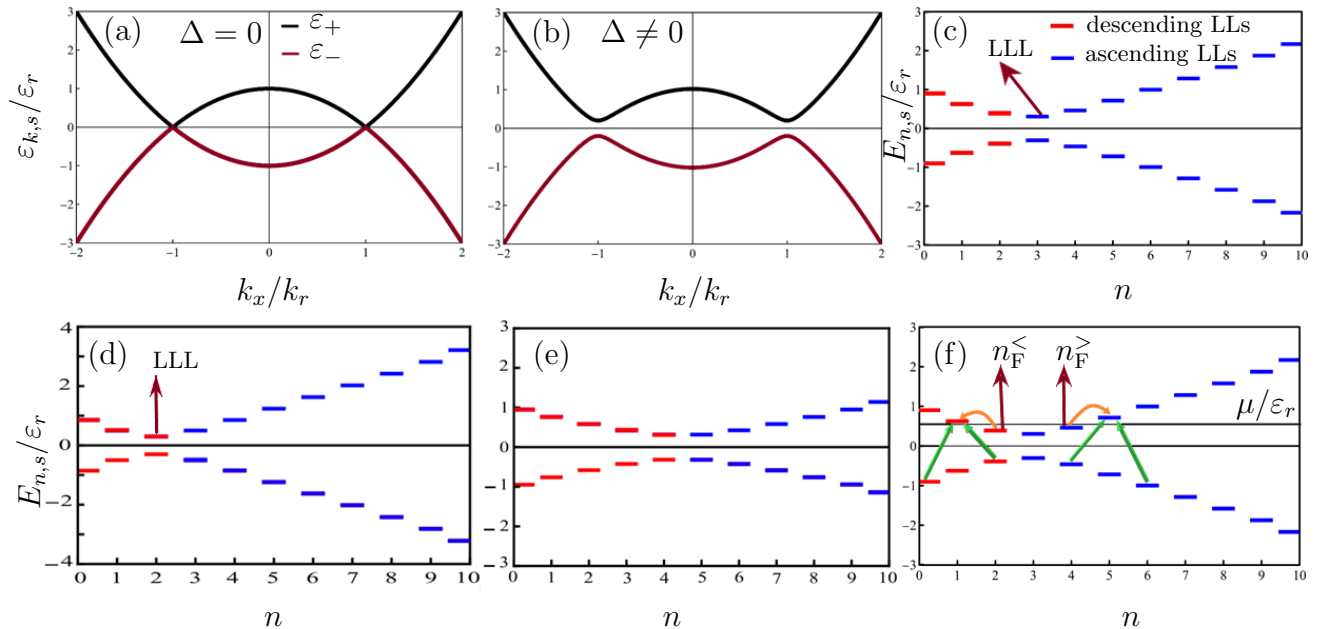

Starting from a generic two-band model Hamiltonian
\be
{\cal H}(\kv)=\hv(\kv)\cdot\tauv,
\ee
where $\hv(\kv)=(h_x(\kv),h_y(\kv),h_z(\kv))$ is a vector function of the 2D wave vector $\kv=(k_x,k_y)$, and $\tauv=(\tau_x,\tau_y,\tau_z)$ is a vector of the Pauli matrices that act on a pseudo-spin (i.e., sublattice or orbital) degree of freedom,
we can model an isotropic 2D NLSM with  $h_x(\kv)=\hbar^2(k^2-k_r^2)/(2m^*)$, and $h_y=h_z=0$~\cite{ekstrom2021kerr,martin2018parity}. 
Here $m^*$ is the band mass, $k=\sqrt{k^2_x+k^2_y}$, and $k_r$ is the radius of the nodal ring that forms at the zero energy~\cite{li2016topological,shahin_prb2017,PhysRevB.109.045120}. The energy dispersion of isotropic 2D NLSM is depicted in panel  (a)  of Fig.~\ref{fig:LL}.
Hybridization of two bands, e.g., due to the spin-orbit coupling, opens up a finite gap between two bands. We can include such a gap in the Hamiltonian through $h_z=\Delta$~\cite{lu2017two,li2016topological}. The gapped two-band model is no longer a semimetal. However, for small $\Delta$, the energy spectrum still resembles the nodal-line-like structure with energy extremes at  $k=0$ and $k=k_r$ [see, panel  (b)  of Fig.~\ref{fig:LL} for an illustration].

Applying a uniform magnetic field $\Bv = B \hat{z}$ perpendicular to the gapped 2D NLSM (we take the $z$ direction normal to the material plane), we can incorporate the orbital effects of the magnetic field in our Hamiltonian via the minimal coupling $\pv \to \pv+e \Av$, to find
\be\label{eq:HamilNLSM_B}   
{\cal H}=\left[\frac{(\pv+e \Av)^2}{2m^*}-\varepsilon_{r}\right]{\tau}_{x}+\Delta \tau_z,
\ee
where $\pv$ is the momentum operator, $\Av$ is the vector potential such that $\Bv=\nabla \times \Av$, $ -e<0 $ is the charge of electron, and $\varepsilon_r=\hbar^2k_r^2/(2m^*)$. 
Choosing to work with the Landau gauge $\Av=B(-y, 0)$, that preserves the translational invariance along the $x$-direction, we can introduce the ladder operators as
 \be\label{eq:ladder}
 \begin{split}
a &=\dfrac{1}{\sqrt{2}}(\xi +\partial_{\xi}),\\
a^\dagger&=\dfrac{1}{\sqrt{2}}(\xi -\partial_{\xi}),
\end{split}
\ee
where $\xi$ is the dimensionless center of the Landau orbits $\xi= y/\ell_B-\ell_B k_x$, with $\ell_B=\sqrt{\hbar/(eB)}$. 
In terms of the ladder operators $a$ and $a^\dagger$, the Hamiltonian~\eqref{eq:HamilNLSM_B} takes the following form 
 \be\label{eq:Hamil2Dn}
{\cal H}=\left[\varepsilon_c(a^\dagger a+\dfrac{1}{2})-\varepsilon_r \right]{\tau}_{x}+\Delta {\tau}_{z}.
\ee
Here, $\varepsilon_c = \hbar \omega_c$  is the cyclotron energy, with $\omega_c= eB/m^*$ the effective cyclotron frequency.

Solving the eigenvalue problem for the Hamiltonian~\eqref{eq:Hamil2Dn}, we find the Landau levels (LLs) as
$E_{n,s} = s E_n$, where $s=+ (-)$ labels the electron (hole) bands, and
\be\label{eq:LL_energy}
E_n= \sqrt{(\varepsilon_n -\varepsilon_r)^2+\Delta^2 },\quad n=0,1,2,...~,
\ee
with $\varepsilon_n=\varepsilon_c(n+1/2)$.
For the Landau gauge we have chosen for the vector potential, the Hamiltonian of Eq.~\eqref{eq:HamilNLSM_B}  commutes with $p_x$, and the energy eigenfunctions read
\be\label{eq:psi}
\psi_{n,k_x,s}(\rv)=\dfrac{e^{ik_x x}}{\sqrt{2L_x}}
\left(
\begin{array}{c}
A^{s+}_n\\
 s \sgn(\sin\theta_n) A^{s-}_n 
\end{array} \right) \varphi _{n}(\xi).
\ee
Here, $ L_x $ is the size of the system in the $ x $ direction, $\sgn(x) $ is the sign function, $ A^{s\pm}_n=\sqrt{1\pm s~\cos\theta_n} $, $ \theta_n=\arctan [(\varepsilon_n-\varepsilon_r)/\Delta]$, and finally\be
\varphi_n(\xi)=\dfrac{1}{\sqrt{2^{n}n!}}\left(\dfrac{m^*\omega_c}{\pi\hbar}\right)^{1/4}e^{-\xi^2/2}H_{n}(\xi),
\ee
is the wave function of the quantum harmonic oscillator, given in terms of $n$-th Hermite polynomial $H_{n}(\xi)$.

We can also define the current-density operator as ${\bf j}= -e {\bf v} $, where ${\bf v} $ is the velocity operator. Using the Heisenberg equation of motion $ {\bf v}={\dot{\rv}}=[\rv, H]/i\hbar $,  the components of the current-density vector of our system read
 \be\label{eq:currentcomp}
 \begin{split}
 	j_{x}&=\dfrac{e}{m^*}\sqrt{\hbar e B }\xi \tau_x =j_c(a+a^\dagger)\tau_x , \\
 	j_{y}&=i\dfrac{e}{m^*}\sqrt{\hbar e B } \partial_{\xi} \tau_x =i j_c (a-a^\dagger) \tau_x .
 \end{split}
 \ee 
Here, $ j_c=\sqrt{e^2\varepsilon_c/(2m^*)} $ and the second equalities on the right-hand-sides follow from the definition of ladder operators in Eq.~\eqref{eq:ladder}. 
%%%%%%%%%%%%Fig. 1%%%%%%%%%%%%%%%%%%%%

%%%%%%%%%%%%%%%%%%%%%%%%%%%
\subsection{Classifications of the Landau levels and different carrier doping regimes \label{sec:classification}}
Looking back at the energies of the Landau levels in Eq.~\eqref{eq:LL_energy}, we notice that depending on the sign of $\varepsilon_n-\varepsilon_r$, we have two sets of Landau levels.
For $\varepsilon_n<\varepsilon_r$, the energy $E_n$ decreases by the LL index $n$, while it increases for $\varepsilon_n>\varepsilon_r$. 
Based on this, we call the LLs with $n\leq \varepsilon_r/\varepsilon_c-1/2$ ($n>\varepsilon_r/\varepsilon_c-1/2$) descending (ascending) Landau levels [see panel (c)-(e)  in Fig.~\ref{fig:LL}]. 
Notice that we use this naming convention based on the absolute value of the LL energy (i.e., $E_n$). Therefore, both conduction and valance bands are categorized in the same manner.  
We also define the index of the lowest Landau level (LLL) as $n_{\rm LLL}={\rm nint}(\varepsilon_r/\varepsilon_c-1/2)$, where ${\rm nint}(x)$ rounds $x$ to the nearest integer number (half-integers are rounded down).  
The LLL can belong to either of the descending or ascending sets of LLs depending on the ratio $\varepsilon_r/\varepsilon_c$ [see $n_{\rm LLL}$ in panels (c) and (d) of Fig.~\ref{fig:LL}].
Notice that two special arrangements for the LLs require extra consideration. First, if the magnetic field is tuned such that $\varepsilon_r/\varepsilon_c$ is half-integer, then $E_{n_{\rm LLL}}=\Delta$, and we have $E_{n_{\rm LLL}-j}=E_{n_{\rm LLL}+j}$ (with $j=1,...,n_{\rm LLL}$), i.e., descending and ascending LLs are degenerate.
Another interesting arrangement happens for integer $\varepsilon_r/\varepsilon_c$. In this case, again, we have degeneracy between descending and ascending LLs, $E_{n_{\rm LLL}-j+1}=E_{n_{\rm LLL}+j}$ ($j=1,...,n_{\rm LLL}$). The only difference with the previous case is the degeneracy of LLL itself, i.e., $E_{n_{\rm LLL}}=E_{n_{\rm LLL}+1}=\sqrt{\varepsilon^2_c/4+\Delta^2}$. 
In panel (c) of Fig.~\ref{fig:LL}, we show the Landau level structure of a  2D NLSM for a generic magnetic field. Descending and ascending LLs, as well as the LLL, are identified in the figure. Panels (d) and (e) of Fig.~\ref{fig:LL} show two special arrangements of the LLs, corresponding to the half-integer and integer  $\varepsilon_r/\varepsilon_c$, respectively.
When the system is doped with the chemical potential $\mu$ (just for the sake of definiteness, we consider electron-doped cases here, but our results for the conductivity are independent of $\mu$'s sign), we identify the highest occupied level, from the descending set of LLs, as 
\be
n^<_\mathrm{F}=
\left\{
\begin{matrix}
	{\rm nint}\left[\left(\varepsilon_r-\sqrt{\mu^2-\Delta^2}\right)/\varepsilon_c\right], & E_{n_{\rm LLL}}<\mu<\varepsilon_r\\
	0, &  \mu>\varepsilon_r
\end{matrix}\right.,
\ee
and, similarly, for the highest occupied LL from the ascending set as
\be\label{eq:nlandau1}
	n^>_\mathrm{F}={\rm nint}\left[\left(\varepsilon_r+\sqrt{\mu^2-\Delta^2}\right)/\varepsilon_c\right]-1,\quad E_{n_{\rm LLL}}<\mu ,
\ee
and for $0<\mu<E_{n_{\rm LLL}}$, the system is un-doped, i.e., all the LLs from the conduction band are empty. 
Note that doped electrons are only added to the localized orbits in this regime. As the lowest Landau level remains unoccupied, we refer to it as the undoped regime.
Both $n^>_\mathrm{F} $  and $ n^<_\mathrm{F} $ are identified in the schematic illustration of panel $ (f) $ in  Fig.~\ref{fig:LL}.
The total number of filled LLs (in the conduction band) is $n^>_\mathrm{F}-n^<_\mathrm{F}+1$.
Now, we can identify different doping regimes: i) the un-doped regime, where no LL (from the conduction band) is filled, ii) the low-doped regime with $E_{n_{\rm LLL}}<\mu<E_0$, and iii) high-doped regime $\mu>E_0$, where all the descending LLs are occupied.
 
%%%%%%%%%%%%%%%%%%%%%%%%%%%%%%%%%%%%%
\section{Kubo formula for optical conductivity}\label{sec:kubo}
The optical conductivity tensor of charge carriers in an electronic system $\sigma_{\alpha\beta}(\omega)$ ($ \alpha, \beta= x, y$ are the spatial directions) relates the current-density induced in the sample ${\bf J}(\omega)$ to the applied external electric field $\bf{E}(\omega)$. 
Within the linear response theory, the Kubo formula gives the optical conductivity~\cite{Kubo_JPSJ1957,Allen_CCCMS2006}
\be \label{eq:kubo}
\begin{split}
 \sigma_{\alpha \beta}(\omega)= \frac{\hbar g_s}{iS}\sum_{k_x} &\sum_{n,s }\sum_{ m,s'}
\frac{f(E_{n,s})-f(E_{m,s'})}{E_{n,s}-E_{m,s'}}\\
&\times \frac{ j^{n,s ;m,s' }_{\alpha}j^{m,s';n,s}_{\beta}}{\hbar\omega +E_{n,s}-E_{m,s'}+i\eta},
 \end{split}
\ee
where $g_s$ is the band degeneracy factor (due to spin, etc.), $S$ is the sample area, $j^{n,s;m,s'}_{\alpha}=\langle n,s | j_{\alpha}|m,s'\rangle$ are the matrix elements of the current-density operator~\eqref{eq:current} along the $\alpha$-direction, $f(\varepsilon)=1/[e^{\beta(\varepsilon-\mu)}+1]$ is the Fermi-Dirac distribution function (with inverse temperature $\beta=1/(k_{\rm B}T)$ and chemical potential $\mu$), and finally  $\eta=\hbar/\tau$ accounts for the lifetime $\tau$ of energy states. It depends on the relevant scattering mechanism, but we will treat it as a constant phenomenological parameter. 

It is straightforward to evaluate the  matrix elements of the current-density operators to find
\be\label{eq:current}
j^{n,s;m,s'}_{\alpha}j^{m,s';n,s}_{\beta}=\dfrac{1}{2}j_c^2 I^{ss'}_{n,m} \Lambda^{\alpha,\beta}_{n,m},
\ee
where
\be\label{eq:factor1}
I^{ss'}_{n,m}=1-ss' cos(\theta_n+\theta_m),
\ee
is the form factor, and 
\be\label{eq:factor01}
\begin{split}
\Lambda^{x,x}_{n,m}=n\delta_{m,n-1} + (n+1)\delta_{m,n+1},\\
\Lambda^{x,y}_{n,m}=i\left[n\delta_{m,n-1}-(n+1)\delta_{m,n+1}\right],
\end{split}
\ee
enforce the selection rules $m=n \pm 1$ for optical transitions between Landau levels.

In what follows, we will concentrate on the zero temperature (i.e., $T=0$) and clean system (i.e., $\eta\to 0^+$) limits so that we can simplify the conductivity tensor Eq.~\eqref{eq:kubo} to 
\be\label{eq:kubo_ab}
\begin{split}
 \sigma_{\alpha\beta}(\omega)=-i\gamma &\sum_{n,s}\sum_{m,s'}
 \frac{I^{ss'}_{n,m}\Lambda^{\alpha,\beta}_{n,m}}{E_{n,s}-E_{m,s'}}\\
&\times\frac{\Theta(\mu-{E}_{n,s})-\Theta(\mu-{E}_{m,s'}) }{\hbar{\omega} +E_{n,s}-E_{m,s'}+i0^+},
 \end{split}
\ee
with $ \gamma=g_s\sigma_0\varepsilon^2_c/4$, where $\sigma_0=e^2/h$ is the quantum of conductance. 
Also note that the summation over $k_x$ is replaced with the degeneracy factor of LLs, i.e., $\sum_{k_x}1=S/(2\pi \ell_B^2)$, as neither the LL energies nor the matrix elements of the current-density operators depend on $k_x$.

Now, we present our analytical results for the real and imaginary parts of longitudinal conductivities of a gapped 2D NLSM, postponing the discussion of Hall conductivity to the subsequent subsection. 

\subsection{Longitudinal Optical Conductivity}
As our model for the NLSM is isotropic, for the longitudinal optical conductivity (LOC), we have $\sigma_{xx}=\sigma_{yy}$.
We can write the LOC as
$	\sigma_{xx}(\omega)=\sigma^{++}_{xx}(\omega)+\sigma^{-+}_{xx}(\omega)$,
where $\sigma^{++}_{xx}$ and $\sigma^{-+}_{xx}$ are the intra-band and inter-band contributions to the LOC. 
Here,  by inter-band, we mean optical transitions from filled LLs of the valence band to empty LLs in the conduction band, while intra-band transitions refer to optical transitions within the LLs of the same bands.
 
At low doping levels, i.e., $E_{n_{\rm LLL}}< \mu < E_0$, we can write the real parts of the intra-band and inter-band contributions to LOC as (we take $\omega>0$, here and in the rest of this paper)
\be\label{eq:re_xx_++}
\begin{split}
	\mathrm{Re}\,\sigma^{++}_{xx}(\omega)=\frac{\pi \gamma}{\hbar^2}
	 \Bigg[& \dfrac{n^<_\mathrm{F}~I^{++}_{n^<_\mathrm{F},n^<_\mathrm{F}-1}}{\Omega^{-}_{n^<_\mathrm{F}-1,n^<_\mathrm{F}}} \delta \left(\left|\Omega^{-}_{n^<_\mathrm{F}-1,n^<_\mathrm{F}}\right|-\omega \right)\\
&\quad- n^<_{\rm F} \longrightarrow (n^>_{\rm F}+1 )\Bigg],
\end{split}
\ee    
and 
\be\label{eq:re_xx_-+}
\begin{split}
\mathrm{Re}\,\sigma^{-+}_{xx}(\omega)
=\frac{\pi \gamma}{\hbar^2} \sum_{E_n>\mu}
\Bigg[&\dfrac{nI^{-+}_{n,n-1}\delta \left( \Omega^+_{n,n-1}  -\omega\right)}{\Omega^+_{n,n-1}}\\
&\quad+n \longrightarrow (n+1)\Bigg],
\end{split}
\ee
respectively, where $\Omega^{\pm}_{n,m}=(E_n\pm E_m)/\hbar$. 
The $E_n>\mu$ condition in Eq.~\eqref{eq:re_xx_-+} simply excludes filled Landau levels (i.e., $n^<_{\rm F}\leq n\leq n^>_{\rm F}$), from the summation.

Similarly, for the imaginary parts of the intra-band and inter-band contributions to LOC at low dopings, we respectively find
\be\label{eq:im_xx_++}
\begin{split}
	\mathrm{Im}\,\sigma^{++}_{xx}(\omega)=\frac{2\gamma\omega}{\hbar^2}
	 \Bigg\{ &\dfrac{n^<_\mathrm{F} I^{++}_{n^<_\mathrm{F},n^<_\mathrm{F}-1}}{\Omega^{-}_{n^<_\mathrm{F}-1,n^<_\mathrm{F}}\left[\omega^2-\left(\Omega^{-}_{n^<_\mathrm{F}-1,n^<_\mathrm{F}}\right)^2\right]}\\
&\quad- n^<_{\rm F} \longrightarrow (n^>_{\rm F}+1 )\Bigg\},
\end{split}
\ee
and
\be\label{eq:im_xx_-+}
\begin{split}
\mathrm{Im}\,\sigma^{-+}_{xx}(\omega)
=\frac{2\gamma\omega }{\hbar^2} \sum_{E_n>\mu}
\Bigg\{ &\dfrac{ n I^{-+}_{n-1,n}}{\Omega^+_{n-1,n}\left[\omega^2-\left(\Omega^+_{n-1,n}\right)^2\right]}\\
&\quad+n \longrightarrow (n+1)\Bigg\}.
  \end{split}
\ee

As we already mentioned in subsection \ref{sec:classification}, for a highly doped system (i.e., $\mu>E_0$), we find $n^<_{\rm F}=0$. In this regime, due to the Pauli blocking, there is no contribution to the conductivity from the descending Landau levels. This is automatically enforced by replacing $n^<_{\rm F}=0$ in Eqs.~\eqref{eq:re_xx_++}-\eqref{eq:im_xx_-+}.
On the other hand, in the un-doped regime, where there is no filled  Landau level in the conduction band, there is no intra-band contribution in LOC, and the real and imaginary parts of inter-band contribution follow from Eqs.~\eqref{eq:re_xx_-+} and \eqref{eq:im_xx_-+} respectively. In this case, the sum should run over all Landau levels of the conduction band.

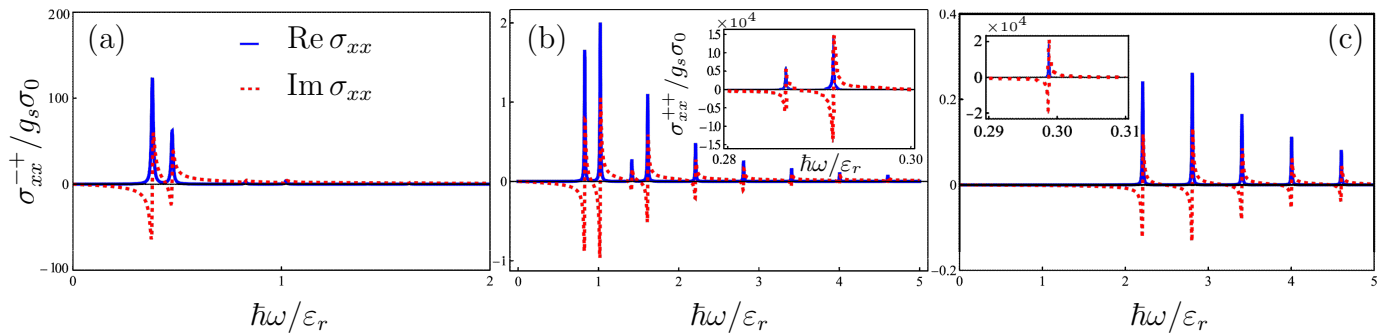
\begin{figure*}    	
	\centering
	\begin{tikzpicture}[scale=0.9]
	\node at (0,0) {\includegraphics[width=\linewidth]{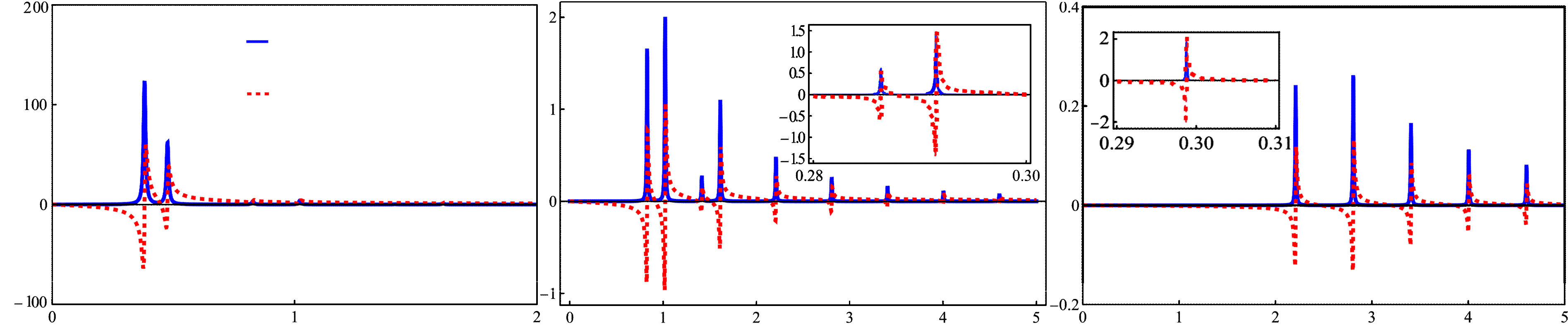}};
	\draw (-5.5,1.6) node [font=\large] {$ \mathrm{Re} \,\sigma_{xx}$};
	\draw (-5.5,0.9) node [font=\large] {$ \mathrm{Im}  \,\sigma_{xx}$};
	\draw (-10,0.1) node [font=\large][rotate=90] {$ \sigma^{-+}_{xx}/g_s \sigma_0 $};
	\draw (-0.4,0.95) node [font=\small][rotate=90] {$ \sigma^{++}_{xx}/g_s \sigma_0 $};
	\draw (-6.1,-2.5) node [font=\large] {$\hbar\omega/\varepsilon_r$};
	\draw (0.4,-2.5) node [font=\large] {$\hbar\omega/\varepsilon_r $};
	\draw (6.9,-2.5) node [font=\large] {$ \hbar\omega/\varepsilon_r $};
	\draw (1.9,-0.25) node [font=\small] {$\hbar\omega/\varepsilon_r$};
	\draw (0.5,1.88) node [font=\tiny] {$\times 10^4$};
	\draw (4.5,1.82) node [font=\tiny] {$\times 10^4$};
	\draw (-8.8,1.6) node [font=\large] { (a) };
	\draw (-2.3,1.6) node [font=\large] { (b) };
	\draw (9.5,1.6) node [font=\large] { (c) };
	\end{tikzpicture}  
   \caption{The real (solid blue lines) and imaginary (red dashed relines) parts of the inter-band longitudinal optical conductivity of a gapped 2D NLSM (in the units of $g_s\sigma_0 $) versus $\hbar\omega/\varepsilon_r$, for a generic case with ${\varepsilon}_r= 3.3\, \varepsilon_c$ in the 
   	(a) un-doped regime with $\mu= 0.1\, \varepsilon_r$;
(b) low doping regime with $\mu= 0.5 \,\varepsilon_r$; and
(c) highly doped system with $\mu= 1.1 \,\varepsilon_r$.
We used $ \Delta=0.1 \,\varepsilon_r $ in all plots and have broadened the Dirac delta functions with parameter $ \eta=5\times 10^{-5}\, \varepsilon_r$ for a better visibility of the peaks. 
Insets: Same as the main panels but for the intra-band contributions to the longitudinal optical conductivity. Note that in the un-doped regime [panel (a)], there is no filled LL and, therefore, no intra-band contribution to longitudinal optical conductivity.
   \label{fig:LOC}}
\end{figure*}

In Fig.~\ref{fig:LOC}, we illustrate the behavior of real and imaginary parts of the longitudinal optical conductivity versus $\omega$ at different doping regimes. 
We notice several interesting features in the LOC's Shubnikov–de Haas oscillations. 
For an un-doped system (see panel (a) in Fig.~\ref{fig:LOC}), all the LLs of the conduction band are empty, so there is no contribution from the intra-band transitions, i.e., $ \sigma^{++}_{xx}(\omega)=0 $.
For a generic magnetic field value and at low doping regimes (as shown in panel (b) of Fig.~\ref{fig:LOC}), two low-frequency double-peaks in the LOC correspond to intra-band transitions
$n^<_{\rm F}\rightarrow n^<_{\rm F}-1$ and $n^>_{\rm F}\rightarrow n^>_{\rm F}+1$. 
The double-peak structure appears in the inter-band transitions for $\omega<2\varepsilon_r/\hbar$ as well, which correspond to inter-band transitions to empty LLs with $n<n^<_{\rm F}$ and $n>n^>_{\rm F}$. 
For $\omega>2\varepsilon_r/\hbar$, there is no contribution from the inter-band transitions to the descending LLs, and single peaks correspond to the inter-band transition to the ascending LLs.  
Note that for some fine-tuned magnetic field strengths, where $\varepsilon_c/\varepsilon_r$ is either integer or half-integer,  the descending and ascending LLs are degenerate, and double peaks at $\omega<2\varepsilon_r/\hbar$ merge.
Panel  (c)  of Fig.~\ref{fig:LOC} shows the behavior of a highly doped system (i.e., $\mu>E_0$), in which all the descending LLs are filled, and the intra-band and all the inter-band transitions arise from the transition to the empty ascending LLs. 

%%%%%%%%%%%%%Fig. 2 %%%%%%%%%%%%%

%%%%%%%%%%%%%%%%%%%%%%%%%%%%%%%%%%
\subsection{Optical Hall conductivity}\label{sec:Hall}
Transverse components of Eq.~\eqref{eq:kubo_ab} give the optical Hall conductivity (OHC). 
We take $\alpha=x$ and $\beta=y$, and in a similar manner to the LOC, we split the OHC into intra-band and inter-band components
$\sigma_{xy}(\omega)=\sigma^{++}_{xy}(\omega)+\sigma^{-+}_{xy}(\omega)$,
where
\be\label{eq:hall8}
\begin{split}
	\mathrm{Re}\,\sigma^{\pm+}_{xy}(\omega)=\frac{2\gamma}{\hbar^2} \Bigg[&\dfrac{n^<_\mathrm{F} I^{\pm+}_{n^<_\mathrm{F}, n^<_\mathrm{F}-1}}{\left(\Omega^{\mp}_{n^<_\mathrm{F}-1,n^<_\mathrm{F}}\right)^2-{\omega}^2}\\
	&- n^<_{\rm F} \longrightarrow (n^>_{\rm F}+1 )\Bigg],
\end{split}
\ee
are the reals parts of the intra-band ($++$) and inter-band ($- +$) contributions to OHC, and their imaginary parts read
\be\label{eq:hall20}
\begin{split}
\mathrm{Im}\,\sigma^{\pm+}_{xy}(\omega)=\frac{\pi \gamma}{\hbar^2}
 \Bigg[& \dfrac{n^<_\mathrm{F}~I^{\pm+}_{n^<_\mathrm{F},n^<_\mathrm{F}-1}}{\Omega^{\mp}_{n^<_\mathrm{F}-1,n^<_\mathrm{F}}}\delta \left(\left|\Omega^{\mp}_{n^<_\mathrm{F}-1,n^<_\mathrm{F}}\right|- \omega \right)\\
 &+ n^<_{\rm F} \longrightarrow (n^>_{\rm F}+1 )\Bigg].
 \end{split}
\ee
Notice that from the optical selection rule of Eq.~\eqref{eq:factor01}, only $n\rightarrow n\pm 1$ transitions are allowed for both longitudinal and Hall responses. 
While the inter-band transitions $n\rightarrow (n+1)$ and $(n+1) \rightarrow n$, with the same resonant energy, add up in the LOC, they cancel each other in the OHC. 
The only inter-band transitions that survive in OHC are $n^<_{\rm F}\rightarrow n^<_{{\rm F}}-1$ and $n^>_{\rm F}\rightarrow n^>_{{\rm F}}+1$ (similar to the allowed intra-band transitions) that their reverse transitions are Pauli blocked. 

Similar to the longitudinal responses, we can find the results for high-doped regime replacing 
$n^<_{\rm F}=0$.
In the un-doped regime, similar to the LOC, there is no intra-band contribution to Hall conductivity. 
Furthermore, inter-band transitions  $n\rightarrow (n+1)$ and $(n+1) \rightarrow n$ cancel out, so an un-doped system's total OHC vanishes.

\begin{figure}
	\centering
	\begin{tikzpicture}[scale=1]
	\node at (0,0) {\includegraphics[width=0.9\linewidth]{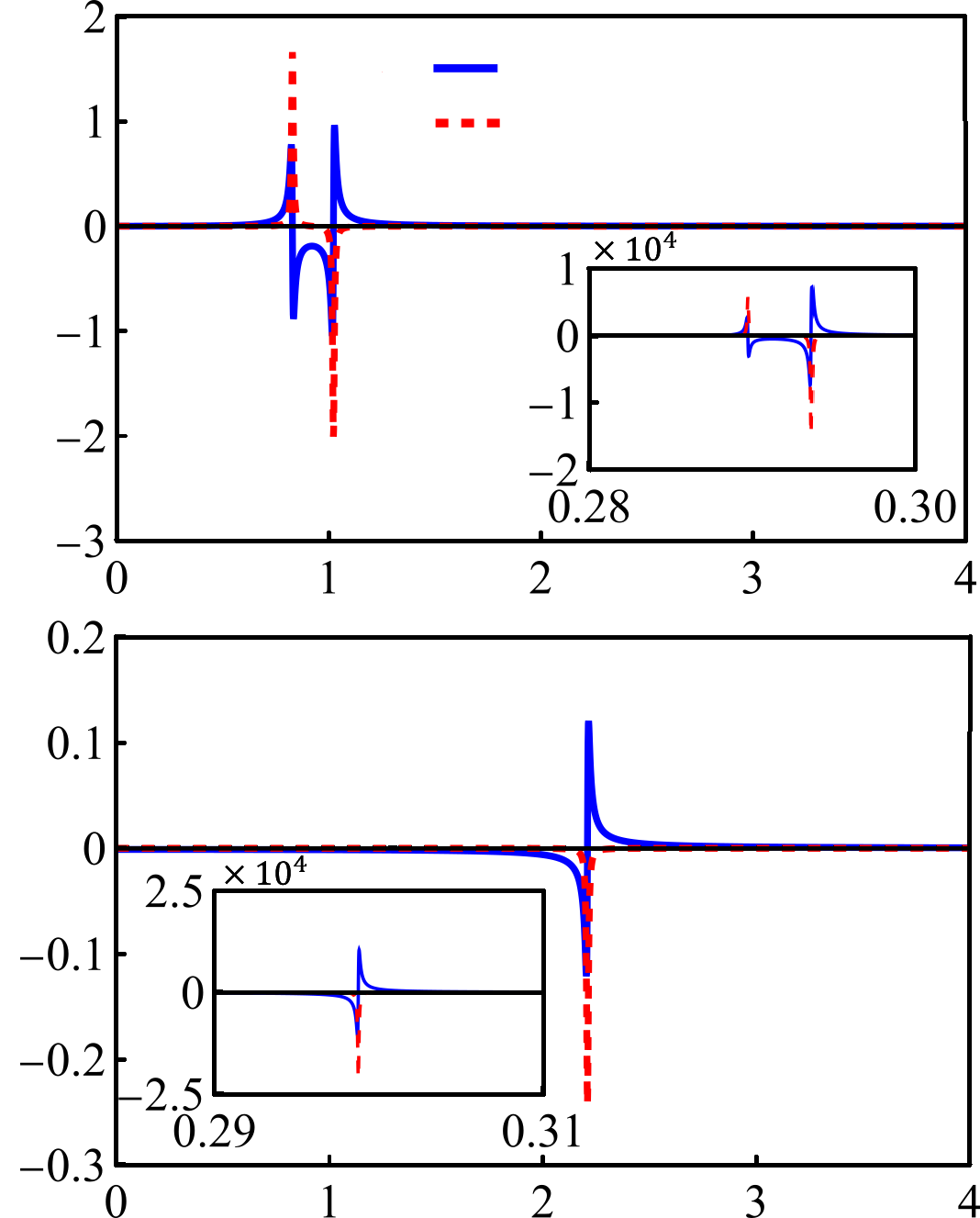}};
	\draw (0.9,3.9) node [font=\large] {$ \mathrm{Im}\, \sigma_{xy} $};
	\draw (0.9,4.4) node [font=\large] {$ \mathrm{Re}\, \sigma_{xy} $};
	\draw (-2.5,4.3) node [font=\large] { (a) };
	\draw (2.4,3.4) node [font=\Large] {$ \mu=0.5\,\varepsilon_r $};
	\draw (-2.5,-0.7) node [font=\large] { (b) };
	\draw (2.4,-1.2) node [font=\Large] {$ \mu=1.1\,\varepsilon_r $};
	\draw (-4.2,3.1) node [font=\large][rotate=90] {$ \sigma^{-+}_{xy}/g_s \sigma_0 $};
		\draw (-4.2,-2.3) node [font=\large][rotate=90] {$ \sigma^{-+}_{xy}/g_s \sigma_0 $};
	\draw (0,2) node [font=\large][rotate=90] {$ \sigma^{++}_{xy}/g_s \sigma_0 $};
	\draw (2.3,0.8) node [font=\large] {$\hbar \omega/\varepsilon_r$};
	\draw (0.7,-5.4) node [font=\large] {$\hbar \omega/\varepsilon_r$};
\end{tikzpicture}  
	\caption{Real (blue solid lines) and imaginary (red dashed lines) parts of the optical Hall conductivity (in the units of  $  g_s\sigma_0 $) versus $ \hbar\omega/\varepsilon_r$, for a generic case with ${\varepsilon}_r= 3.3\, \varepsilon_c$ in the  
(a) low doping regime with $\mu= 0.5 \,\varepsilon_r$;
and (b) high doping regime with $\mu= 1.1 \,\varepsilon_r$.
We have used $\Delta=0.1\, \varepsilon_r$,  and $ \eta=5\times 10^{-5}\, \varepsilon_r$  in all plots. 
Insets: Same as the main panels but for the intra-band contributions to the optical Hall conductivity. 
		\label{fig:OHC}}
\end{figure}

In Fig.~\ref{fig:OHC}, we illustrate the behavior of real and imaginary parts of OHC versus $ \omega $ in the low and high-doped regimes. 
Double peaks at low doping correspond to different inter-band and intra-band transitions. In contrast, for a highly doped system, inter-band and intra-band contributions each contain a single transition peak.
%%%%%%%%%%%Fig. 3%%%%%%%%%%%%%%%%%

%%%%%%%%%%%%%%%%%%%%%%%%%%%%%%%%%%%%%%%
\section{DC Hall conductivity}\label{sec:DC-Hall}
In this section, we discuss the behavior of static conductivity.
In the static (i.e., $\omega \to 0$) limit, the longitudinal conductivity vanishes, and the Hall conductivity becomes a real function. We can easily find the analytic form of the DC Hall conductivity from the zero-frequency limit of Eq.~\eqref{eq:hall8}
\be\label{eq:rehall1}
\sigma^{dc}_{{xy}}=-g_s\sigma _0{\cal C},
\ee
where the Chern number ${\cal C}=n^>_\mathrm{F}+1-n^<_\mathrm{F}$, is the filling factor, i.e., the number of filled LLs in the conduction band.

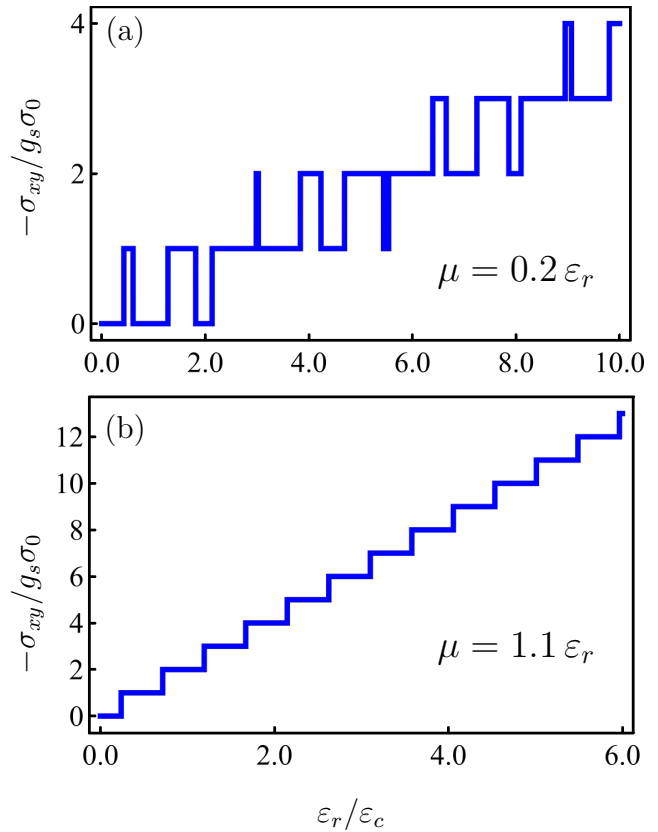
\begin{figure}
\centering
\begin{tikzpicture}[scale=1]
\node at (0,0) {\includegraphics[width=0.9\linewidth]{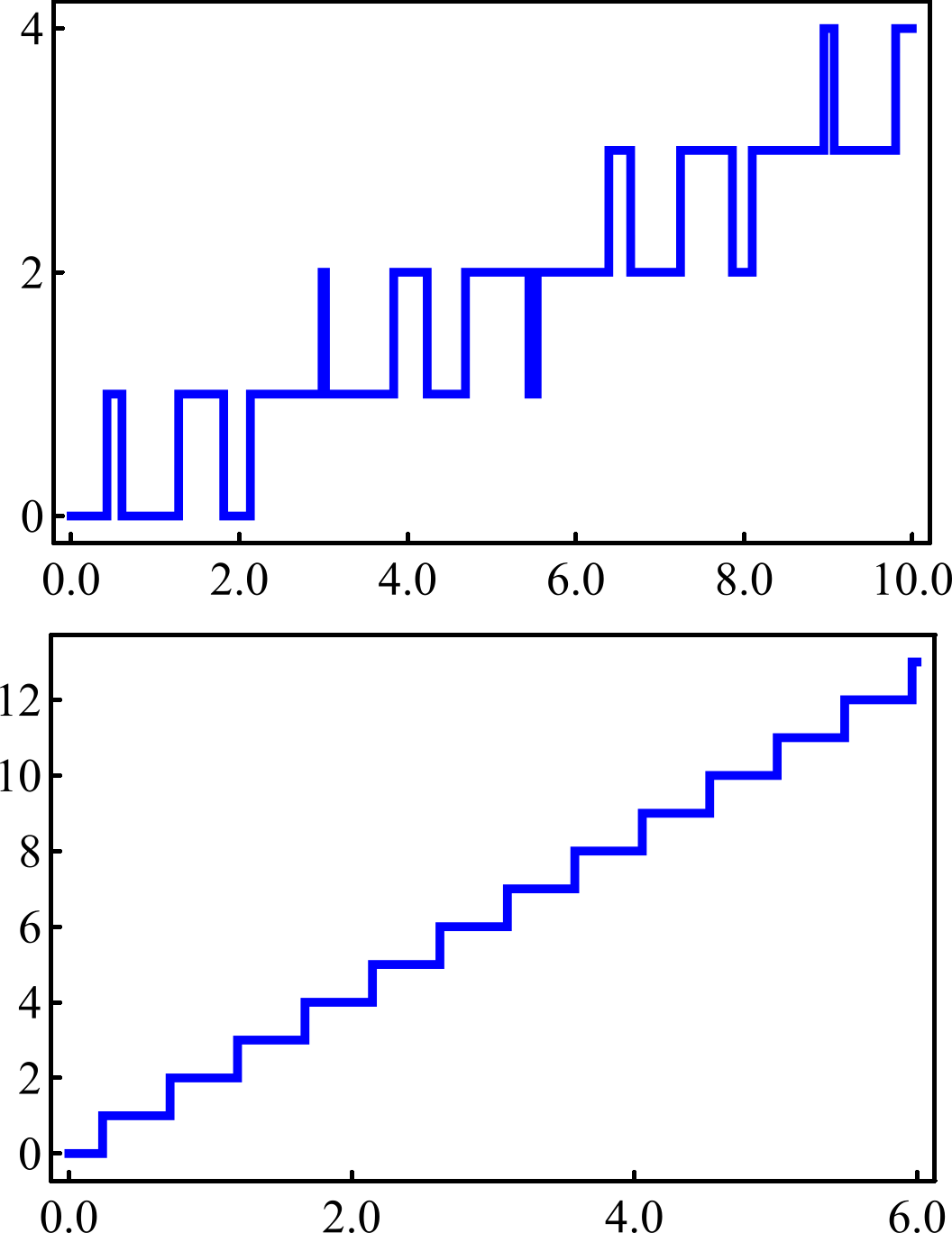} };
\draw (-3,4.7) node [font=\large] {(a)};
\draw (2.2,1.5) node [font=\Large] {$ \mu=0.2\,\varepsilon_r $};
\draw (-3,-0.6) node [font=\large] {(b)};
\draw (2.2,-3.5) node [font=\Large] {$ \mu=1.1\,\varepsilon_r $};
\draw (-4.3,2.9) node [font=\large][rotate=90] {$ -\sigma_{xy}/g_s \sigma_0 $};
\draw (-4.3,-2.9) node [font=\large][rotate=90] {$ -\sigma_{xy}/g_s \sigma_0 $};
\draw (0,-5.7) node [font=\large] {$\varepsilon_r/\varepsilon_c$};
\end{tikzpicture}    
   \caption{ The DC Hall conductivity (in the units of $g_s\sigma_0$) as a function
of the inverse cyclotron energy for 
$ \mu=0.2\,\varepsilon_r $ (a), and
$ \mu=1.1 \,\varepsilon_r $ (b). 
We have fixed the gap parameter in both plots at $\Delta=0.1\,\varepsilon_r$. 
   \label{fig:g1020}}
\end{figure}

We investigate the behavior of static Hall conductivity versus magnetic field at fixed chemical potential values in  Fig.~\ref{fig:g1020}. Fixing the chemical potential while the magnetic field is varied means that the carrier density is not fixed. 
The chemical potential can be fixed experimentally through a gate potential placed on top of the two-dimensional NLSM.
The bumpy Hall plateaus in panel (a) of Fig.~\ref{fig:g1020} originate from the fact that at low doping, as we increase the magnetic field, the descending Landau levels enter below the Fermi level while ascending Landau levels are pushed above it. As the separation between adjacent LLs is not constant, this moving staircase of Landau levels can result in the swinging Chern number.
 In the high doping regime, as illustrated in panel (b) of Fig.~\ref{fig:g1020}, all of the descending LLs are already filled, and we observe only uniform plateaus via changing the magnetic field. The conductivity is quantized in the units of $g_s e^2/h$ in both doping regimes.

\begin{figure}
	\centering
	\begin{tikzpicture}[scale=1]
	\node at (0,0) {\includegraphics[width=0.9\linewidth]{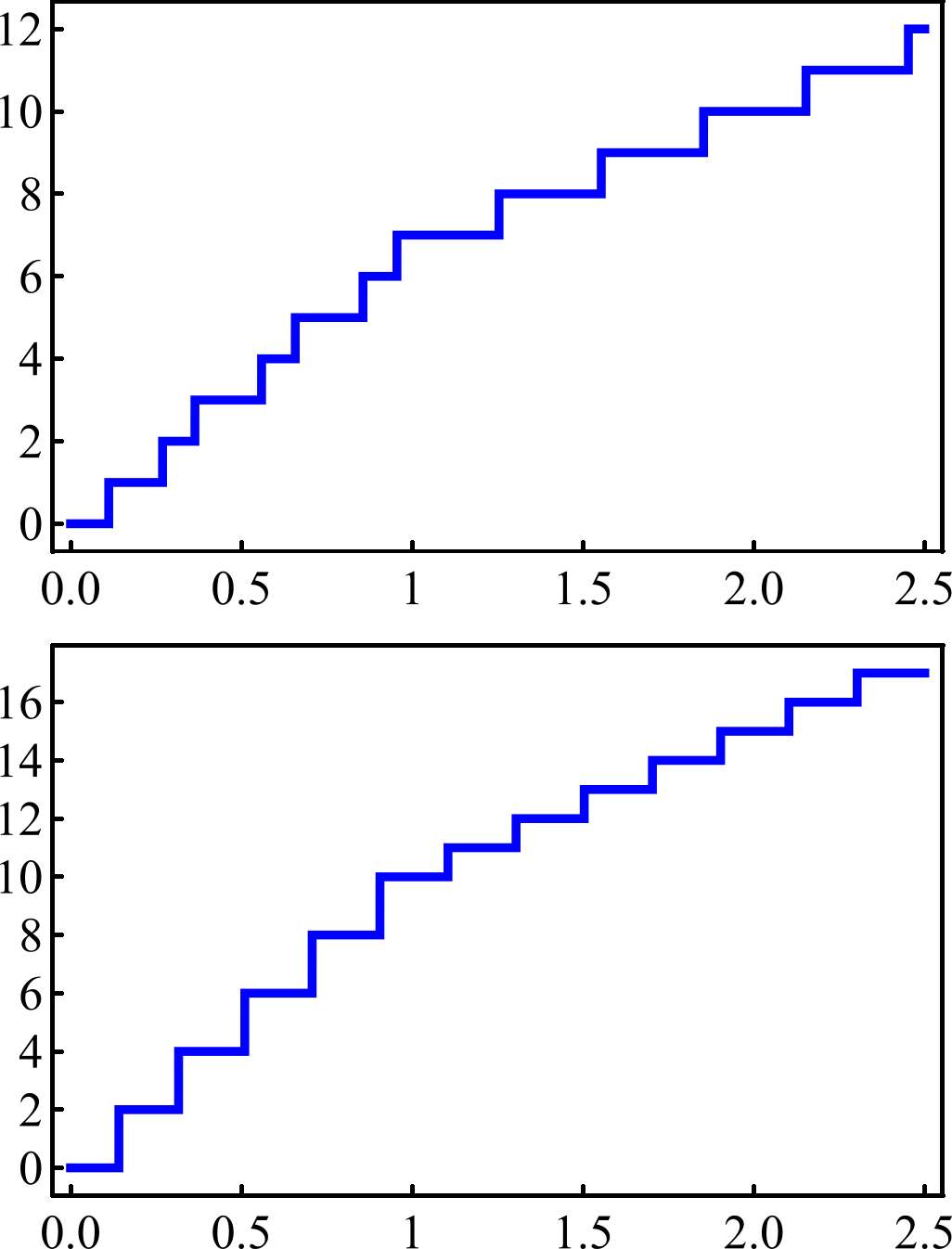} };
	\draw (-3,4.7) node [font=\large] {(a) };
	\draw (2.2,1.5) node [font=\Large] {$ \varepsilon_r=3.3\,\varepsilon_c $};
	\draw (-3,-0.6) node [font=\large] { (b) };
	\draw (2.2,-3.5) node [font=\Large]  {$ \varepsilon_r=5.0\,\varepsilon_c $};
	\draw (-4.6,3.2) node [font=\large][rotate=90] {$ -\sigma_{xy}/g_s \sigma_0 $};
		\draw (-4.6,-2.9) node [font=\large][rotate=90] {$ -\sigma_{xy}/g_s \sigma_0 $};
	\draw (0.3,-6) node [font=\large] {$\mu/\varepsilon_r$};
	\end{tikzpicture}  
	\caption{The DC Hall conductivity (in the units of $g_s \sigma_0 $) as a function
		of the chemical potential (in the units of $\varepsilon_r $) for 
		(a) $ \varepsilon_r= 3.3\,\varepsilon_c $, and
		(b)  $ \varepsilon_r= 5.0\,\varepsilon_c $.
	The gap parameter in both panels is  $\Delta=0.1\, \varepsilon_r$. 
		\label{fig:g100}}
\end{figure}

In Fig.~\ref{fig:g100}, we illustrate the chemical potential dependence of DC Hall conductivity of a 2D NLSM at different magnetic field strengths. 
We look at the general case of the magnetic field in panel (a) of Fig.~\ref{fig:g100}, where the descending and ascending LLs are not degenerate. In this case, we find alternating conductivity plateau sizes for $\mu<\varepsilon_r$, as both descending and ascending LLs get filled one by one through increasing the chemical potential. 
For $\mu>E_0$, all the descending levels are filled, and we are filling the ascending levels by increasing the chemical potential. Therefore, the alternating plateau pattern disappears. 
The plateau sizes are more or less constant for particular magnetic field values where two branches of the LLs are degenerate. However, the jumps in the conductivity are twice the conductance quanta ($g_s\sigma_0$) for $\mu<\varepsilon_r$ as illustrated in panel (b) of Fig.~\ref{fig:g100}.

%%%%%%%%%%%Fig. 4%%%%%%%%%%%%%%%
%%%%%%%%%%%%%%Fig. 5%%%%%%%%%%%%%%%%%%%%%

%%%%%%%%%%%%%%%%%%%%%%%%%
\section{Magneto-conductivity of gapless Nodal-line semimetals}\label{sec:Gapless}
In the absence of hybridization between conduction and valance bands, we have $\Delta=0$. In this limit, two bands touch at zero energy, forming a nodal loop of radius $k_r$ in the $k$-space. 
With an external magnetic field, the energy of Landau levels read $E_n=\left|\varepsilon_n-\varepsilon_r\right|$, and the form factor in Eq.~\eqref{eq:factor1}, becomes $I^{ss'}_{n,m}=1+ss'\sgn{(\varepsilon_n-\varepsilon_r)}\sgn{(\varepsilon_m-\varepsilon_r)}$. This form factor can take only two values, 0 or 2.
Taking the $ \Delta \to 0 $ limit in Eqs.~\eqref{eq:re_xx_++}-\eqref{eq:im_xx_-+}, we find  (all in the units of $g_s\sigma_0$)
\be\label{eq:re_xx_++0}
	\mathrm{Re}\,\sigma^{++}_{xx}(\omega)=\frac{\pi }{2} \omega_c \left(n^>_{\rm F}+n^<_\mathrm{F}+1 \right)  \delta (\omega_c-\omega ),
\ee   
and
\be\label{eq:re_xx_-+0}
	\mathrm{Re}\,\sigma^{-+}_{xx}(\omega)
	= \pi \omega_c    n_\mathrm{LLL}\Theta(E_{n_\mathrm{LLL}}-\mu)
	\delta (\omega_c-\omega),
\ee
for the real parts of the LOC and similarly
\be\label{eq:im_xx_++0}
	\mathrm{Im}\,\sigma^{++}_{xx}(\omega)= \frac{\omega \omega_c }{\omega^2-\omega^2_c}(n^>_{\rm F}+n^<_\mathrm{F}+1 ),
\ee
and
\be\label{eq:im_xx_-+0}
	\mathrm{Im}\,\sigma^{-+}_{xx}(\omega)
	= \frac{2 \omega \omega_c }{\omega^2-\omega^2_c}n_\mathrm{LLL}\Theta(E_{n_\mathrm{LLL}}-\mu),
\ee
for the imaginary part of LOC.
Note that for a gapless system, we have 
\be
n^<_\mathrm{F}=
\left\{
\begin{matrix}
	{\rm nint} \left[\left(\varepsilon_r-\mu\right)/\varepsilon_c \right], & \mu<\varepsilon_r \\
	0, &  \mu>\varepsilon_r
\end{matrix}\right.,
\ee
and
\be\label{eq:nlandau11}
	n^>_\mathrm{F}={\rm nint}\left[\left({\varepsilon_r+\mu}\right)/{\varepsilon_c}\right]-1.
\ee
The double peaks of LOC shown in the insets of Fig.~\ref{fig:LOC} entirely fall on each other as the LLs are equidistant in a gapless system. Note that we have inter-band contribution only in the un-doped regime ($0<\mu<E_{n_{\rm LLL}}$). 
In the special case where $ \varepsilon_r/\varepsilon_c $ is half-integer as we have $E_{n_{\rm LLL}}=0$, the system is doped for $\mu>0$. For integer values of $ \varepsilon_r/\varepsilon_c $, we should substitute $ n_\mathrm{LLL}\to n_\mathrm{LLL}+1 $ in Eqs.~\eqref{eq:re_xx_-+0} and \eqref{eq:im_xx_-+0}.

In a similar fashion to LOC, the  real  and imaginary parts of the OHC (again, in the units of $g_s\sigma_0$) read
\be\label{eq:re_xy_++0}
	\mathrm{Re}\,\sigma^{++}_{xy}(\omega)=-\frac{\omega_c^2}{\omega^2_c-\omega^2} {\cal C} ,
\ee    
and 
\be\label{eq:Im_xy_++0}
	\mathrm{Im}\,\sigma^{++}_{xy}(\omega)=-\frac{\pi \omega_c}{2} {\cal C} \delta (\omega_c-\omega ).
\ee     
Note that the inter-band contribution to the Hall conductivity in the gapless limit vanishes.
In the static limit, we arrive again at Eq.~\eqref{eq:rehall1} for the DC Hall conductivity in terms of the filling factor ${\cal C}$.

%%%%%%%%%%%%%%%%%%%%%%%%%
\section{Summary and Conclusion}\label{sec:summary}
We have studied gapped 2D NLSMs in a perpendicular magnetic field. The Landau levels of our model exhibit a non-monotonic behavior forming a \emph{stretched check-mark} shape, where the energies of levels decrease versus the level index up to the lowest Landau level, then start increasing with increasing the level index.

Increasing the strength of the magnetic field at low chemical potentials pushes the descending LLs below the chemical potential while ascending levels rise above it and get vacant. 
Due to this remarkable behavior, we observe that the filling factor (i.e., the Chern number) and, therefore, the static Hall conductivity swings back and forth between different values as we change the magnetic field at a fixed chemical potential. 

We also investigate the conductivities in the intrinsic and different doping regimes. 
The filling factor is zero in an intrinsic (un-doped) gapped 2D NLSM, and the optical and DC Hall conductivities vanish. Furthermore, only the inter-band transitions contribute to the longitudinal optical conductivity.
At low doping, the optical transition among the descending and ascending Landau levels results in a series of double absorption peaks in the longitudinal optical conductivity at low frequencies. 
In the optical Hall conductivity, the inter-band transitions from $ n \rightarrow n+1 $ and $ n+1\rightarrow n $ cancel each other, and only two optical transitions to the first empty levels above the Fermi energy survive, as their reverse process is Pauli blocked. Hence, the inter-band and intra-band contributions to the optical Hall conductivity show a double-peak (single-peak) structure at low (high) doping.

\acknowledgments
SHA thanks M. R. Kolahchi and B. Farnudi for their insightful suggestions.
This work is supported by the Iran Science Elites Federation (ISEF) and the Research Council of the Institute for Advanced Studies in Basic Sciences (IASBS).

%\bibliography{reff}
\bibliography{NLSHall.bbl}

%%%%%%%%%%%%%%%%%%%%%%%%

\end{document}